\documentclass[acmsmall, screen=true, nonacm]{acmart}
\usepackage[ruled,linesnumbered]{algorithm2e}
\usepackage{xcolor}
\usepackage{algpseudocode}
\usepackage{ebproof}
\usepackage{multirow}
\usepackage{tabularx}
\usepackage{bm}
\usepackage{diagbox}
\usepackage{multicol}
\usepackage{colortbl}
\usepackage{xcolor}
\usepackage{makecell}
\usepackage[export]{adjustbox} 
\usepackage{amsmath}
\usepackage{ebproof} 
\usepackage[T1]{fontenc}
\usepackage{upquote}
\usepackage{wrapfig}
\usepackage[most]{tcolorbox}
\usepackage{threeparttable} 
\usepackage{graphics}
\usepackage{epsfig}
\usepackage{setspace}
\usepackage{mathtools}
\usepackage{mathrsfs}
\usepackage{mathpartir}
\usepackage{xspace}
\usepackage{extarrows}
\usepackage{cancel}
\usepackage{tabu}
\usepackage{graphicx}
\usepackage{subcaption}
\usepackage{color}
\usepackage{booktabs}
\definecolor{darkgreen}{rgb}{0,0.6,0}
\definecolor{keywordcolor}{rgb}{0.8,0.1,0.5}
\usepackage{listings}
\usepackage{tikz}
\usepackage{pifont}% http://ctan.org/pkg/pifont
\newcommand{\tit}[1]{\textit{#1}}
\newcommand*{\circled}[1]{\lower.7ex\hbox{\tikz\draw (0pt, 0pt)%
    circle (.4em) node {\makebox[1em][c]{\small #1}};}}
\usepackage[inline,shortlabels]{enumitem}
\usepackage{eucal} % change the font of \mathcal
\newcommand{\bluec}[1]{{\color[rgb]{0,0,0.8}#1}}

\newcommand{\var}[1]{{\color[rgb]{0.6,0,0}#1}}

\newcommand{\mainname}{\textsc{LGuess}\xspace}
\newcommand{\myautoref}[2]{\hyperref[#2]{#1~\ref*{#2}}}

\newcommand{\toolname}[1]{\textsc{#1}}
\makeatletter
\renewcommand\paragraph{\@startsection{paragraph}{4}{0pt}%
  {-.2\baselineskip \@plus -2\p@ \@minus -.2\p@}%
  {-3.5\p@}%
  {\ACM@NRadjust{\bfseries\itshape\@adddotafter}}}
\makeatother

\newcommand{\ruyidel}[1]{}

\ifx\formal\undefined
  
\else
   
\fi

\newcommand{\bb}{\begin{array}{llllllll}}
\newcommand{\ee}{\end{array}}
\SetFuncSty{textsf}
\SetArgSty{textnormal}

\hypersetup{colorlinks,
  linkcolor=ACMDarkBlue,
  citecolor=ACMPurple,
  urlcolor=ACMDarkBlue,
  filecolor=ACMDarkBlue}
\usepackage{booktabs}   %% For formal tables:
\title{Equality Saturation Guided by Large Language Models}

\author{Wentao Peng}
\affiliation{
  \streetaddress{Key Lab of High Confidence Software Technologies, Ministry of Education Department of Computer Science and Technology, School of Computer Science}
  \institution{Peking University}
  \city{Beijing}
  \country{China}            
}
\authornote{Equal contribution}
\email{2200012940@stu.pku.edu.cn}

  \author{Ruyi Ji}
  \authornotemark[1]
\affiliation{
  \streetaddress{Key Lab of High Confidence Software Technologies, Ministry of Education Department of Computer Science and Technology, School of Computer Science}
  \institution{Peking University}
  \city{Beijing}
  \country{China}            
}
\email{jiruyi910387714@pku.edu.cn}

\author{Yingfei Xiong}
\affiliation{
  \streetaddress{Key Lab of High Confidence Software Technologies, Ministry of Education Department of Computer Science and Technology, School of Computer Science}
  \institution{Peking University}
  \city{Beijing}
  \country{China}            
}
\authornote{Corresponding author}
\email{xiongyf@pku.edu.cn}
 
\begin{abstract} 
  One critical issue with LLMs is their inability to guarantee correctness. 
  Although this problem can be addressed by applying LLMs to formal rewrite systems, the capability of LLMs is still far from adequate to generate sound rewrite chains.
  To bridge this gap, this paper proposes \textit{LLM-guided equality saturation}, dubbed as \mainname, by incorporating e-graphs as an intermediate layer between LLMs and rewrite systems. 
  \mainname queries LLMs for only high-level rewrite checkpoints and uses e-graphs to supply low-level rewrite chains between these checkpoints.
  In this procedure, the key technical challenge lies in effectively extracting a suitable checkpoint from a saturated e-graph,
  and \mainname addresses this by learning a probabilistic model from the LLM.
  The model predicts probable checkpoints while remaining simple enough for effective extraction.
  
  We have implemented a prototype of \mainname and evaluated it on the problem of factorizing multi-variable polynomials. 
  The results demonstrate a significant advantage of \mainname compared to both straightforward equality saturation and the approach that queries the LLM directly for the rewrite chain.
\end{abstract}

\begin{document}
\maketitle

%!TEX root = ../paper.tex
\section{Introduction} \label{section:intro}
Recently, LLMs have revealed remarkable effectiveness in program optimization.
They have been applied to optimize programs in different scenarios, such as compile-time optimization~\cite{lange2025ai, DBLP:conf/cc/CumminsSGRGSL25}, competitive programming~\cite{DBLP:journals/tse/ChenFM24, DBLP:journals/corr/abs-2408-12159}, and software development~\cite{DBLP:conf/sigsoft/GargMCSW22, DBLP:journals/corr/abs-2306-17077},
achieving impressive results that sometimes overwhelm traditional techniques, especially on complex tasks that require large-scale modifications.

Despite these achievements, correctness remains a serious issue of LLM-based optimizers. 
LLMs fundamentally cannot ensure the correctness of their output, making it dangerous to adopt their optimizations.
For example, \citet{lange2025ai} report that their LLM-based optimizer can produce incorrect results even against a testing script -- the optimizer discovers a memory exploit in the testing script and sometimes utilizes the exploit to escape the correctness test.

\paragraph{Rewrite systems} One promising method for ensuring correctness is to \textit{certify the optimizations from LLMs through formal rewrite systems}.
Rather than directly generating the optimized program, we can apply LLMs to a pre-defined rewrite system and produce a step-by-step rewrite chain for the optimization.
In this way, correctness is guaranteed by the rewrite system, and the result can be formally verified by confirming that each step soundly applies a rewrite rule.

However, the current capability of LLMs is still far from adequate to generate rewrite chains.
\begin{itemize}
    \item On the one hand, modern rewrite systems are too complex for LLMs to fully grasp.
    For example, \toolname{Szalinski}~\cite{DBLP:conf/pldi/NandiWAWDGT20}, a tool for optimizing CAD programs, implements 65 rules across \textasciitilde1,000 LOC and incorporates custom solvers for arithmetic reasoning and list partitioning -- 
    such a large system cannot even be fully described in a prompt.
    \item On the other hand, most rewrite systems operate at a low level, making the rewrite chains too intricate for LLMs to maintain.
    For example, \toolname{Szalinski} often applies hundreds of rules to optimize CAD programs, resulting in complex rewrite chains comprising millions of AST nodes --
    this scale far exceeds the capabilities of current LLMs.
\end{itemize}

\paragraph{Our approach} To bridge the gap, this paper proposes \textit{LLM-guided equality saturation}, dubbed as \mainname, by incorporating e-graphs as an intermediate layer between LLMs and rewrite systems.

\mainname is motivated by the recent work on \emph{guided equality saturation}~\cite{DBLP:journals/pacmpl/KoehlerGBGTS24}, which demonstrates that human users can be aware of important checkpoints in the rewrite chain even without knowing details on the rewrite system.
Based on this observation, \mainname applies LLMs to assume the role of human users in suggesting useful checkpoints, as LLMs have exhibited human-like capabilities; and then utilizes e-graphs to supply the rewrite chains between the checkpoints, mostly low-level rewrites that cannot be identified without knowing the rewrite system.
 
\begin{figure*}[t]
    \begin{center}
        \includegraphics[width=0.8\textwidth]{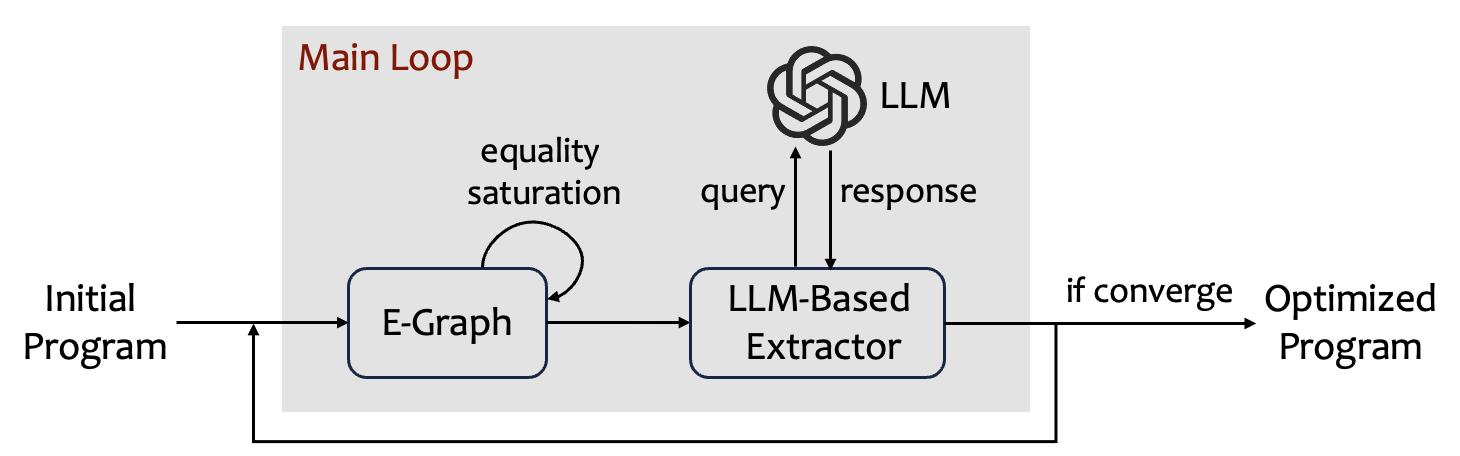}
        \vspace{-0.3em}
        \caption{The workflow of \mainname.}
        \vspace{-1.5em}
        \label{figure:overview} 
    \end{center} 
\end{figure*}

\autoref{figure:overview} illustrates the workflow of \mainname, which runs in multiple rewrite phases.
In each phase, \mainname creates a new e-graph for the current program and applies equality saturation until reaching a resource limit.
Then, \mainname queries LLMs to identify a checkpoint program from the e-graph and takes the checkpoint as the input of the next phase.
This cycle repeats until convergence, i.e., when the checkpoint program remains unchanged after one phase.

\paragraph{Extraction from e-graphs} A key technical problem here is how to effectively extract a suitable checkpoint from the saturated e-graph.
After equality saturation, an e-graph is typically huge in its size and the number of represented programs.
Hence, it is impractical to describe the entire e-graph in a prompt or to query the LLM individually for each candidate program.

\mainname solves this problem by learning a probabilistic model from the LLM, which predicts probable checkpoints and remains simple enough to enable efficient extraction.
\mainname extracts from the e-graph by iteratively refining a default model.
In each round, it samples a new program according to the current model and queries the LLM to compare the new program and the previous result.
Given the response, \mainname refines the model by enhancing the better program and weakening the worse one, and updates the result if the new program is preferred.
After repeating this process for a pre-defined number of rounds, \mainname takes the final result as the checkpoint extracted from the e-graph.

\paragraph{Evaluation} We implement a prototype of \mainname and evaluate it on the problem of factorizing multi-variable polynomials.
This problem is challenging for traditional program rewrite techniques because it requires numerous applications of the associative and commutative laws, which will typically cause a serious combinatorial explosion in the search space.

The results demonstrate the effectiveness of \mainname~--~it has a significant advantage compared to both straightforward equality saturation and querying the LLM directly for the rewrite chain.
%!TEX root = ../paper.tex
\section{Overview} \label{section:overview}
We illustrate \mainname using the following task, which is often dubbed as the \emph{freshman's dream}.
\begin{center}
    \fbox{Simplify the expression $(x+y)^2$ on a commutative ring with characteristic 2.}
\end{center}
The expected result is $x^2 + y^2$. 
\autoref{fig:rules} lists the rewrite rules available in this task, and \autoref{fig:sketch} outlines a rewrite chain for the simplification, which applies $13$ rewrites in total.

\toolname{GPT-4o} can identify the expected result given the above task description, but it fails to construct a sound rewrite chain using the rewrite rules.
On the other hand, this task is not trivial for equality saturation -- it takes \toolname{egg}~\cite{DBLP:journals/pacmpl/WillseyNWFTP21} 7 iterations to find the target expression.

Below, we show how \mainname solves this task under an extreme restriction on equality saturation, where we assume \toolname{egg} can only run up to $2$ iterations when rewriting each program.

\begin{figure*}
    \begin{minipage}{0.45\textwidth}
        \small
        \centering
        \begin{tabular}{lc}
            %\Xhline{1pt}
            %ID & Rule  \\
            %\Xhline{1pt}
            \textsc{(Sqr)} & $a^2 \Leftrightarrow a \cdot a$ \\
            \textsc{(Char-2)} & $a + a \Rightarrow 0$ \\
            \textsc{(Add-0)} & $a + 0 \Rightarrow a$ \\ 
            \textsc{(Add-C)} & $a + b \Leftrightarrow b + a$ \\
            \textsc{(Times-C)} & $a \,\cdot\, b \Leftrightarrow b \,\cdot\, a$ \\ 
            \textsc{(Add-A)} & $a + (b + c) \Leftrightarrow (a + b) + c$ \\
            \textsc{(Distr)} & $(a + b) \cdot c \Rightarrow a \cdot c + b \cdot c$ \\
            %\Xhline{1pt}
           \end{tabular}
        \caption{Rewrite rules in a ring with char. 2.}
        \label{fig:rules}
    \end{minipage}
    \hfill\ 
    \begin{minipage}{0.53\textwidth}
    \small
    \centering
        \begin{tabular}{lll}
            & $(x + y)^2$ \\
            $\Rightarrow^*$ & \quad  $\var{\{\textsc{Sqr}, \textsc{Distr}, \textsc{Times-C} \times 2\}}$\\ 
            & $(x + y) \cdot x + (x + y) \cdot y$ \\
            $\Rightarrow^*$ & \quad $\var{\{\textsc{Distr} \times 2, \textsc{Add-A} \times 2, \textsc{Times-C}\}}$\\ 
            & $(x \cdot x + (x \cdot y + x \cdot y)) + y \cdot y$ \\ 
            $\Rightarrow^*$ & \quad $\var{\{\textsc{Sqr} \times 2, \textsc{Char-2}, \textsc{Add-0}\}}$\\
            & $x^2 + y^2$ 
           \end{tabular}
        \caption{The outline of the simplification.}
        \label{fig:sketch}
    \end{minipage}
\end{figure*}

\paragraph{Walkflow} \mainname runs by phases.
In the first phase, \mainname starts with creating an e-graph for the initial expression $(x + y)^2$ and saturates it within the resource limit.

\begin{figure*}
\begin{minipage}{0.45\textwidth}
    \vspace{0.5em}
    \centering
    \includegraphics[width=0.8\textwidth]{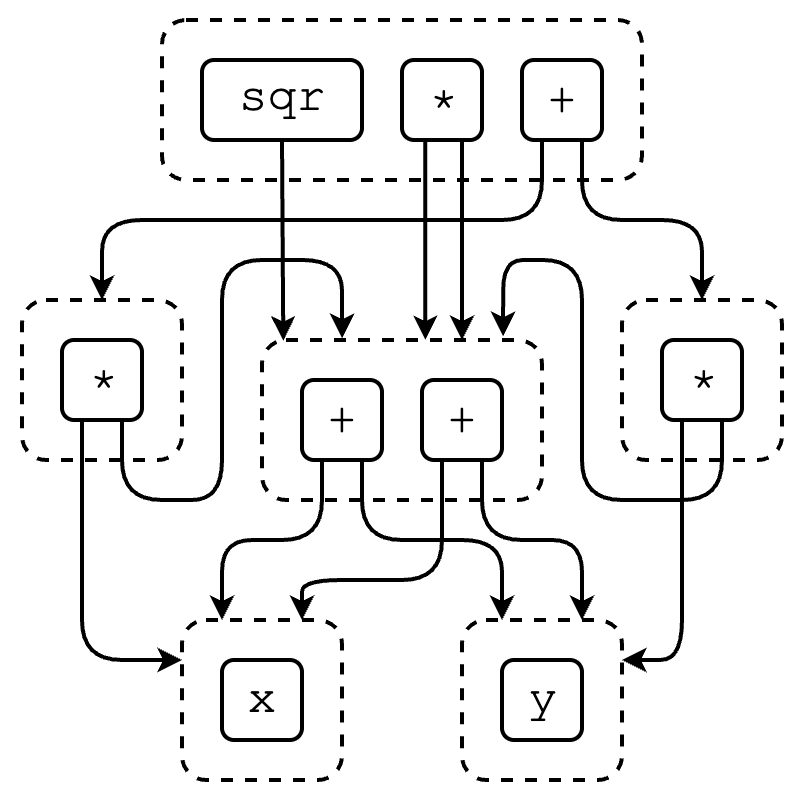}
    \caption{The saturated e-graph in the first phase, where $\tit{sqr}$ denotes the square operator $(\cdot)^2$, and $*$ denotes the multiplication operator.}
    \label{fig:first-egraph}
\end{minipage}
\hfill 
\begin{minipage}{0.5\textwidth}
    \begin{table}[H]
    \caption{Our bigram models, where each row denotes an operator and each column denotes a context.}
    \label{table:models}
    \small 
    \begin{subtable}{\textwidth}
        \vspace{-0.2em}
        \centering
        \caption{The default model.} \label{table:default-model}
        \begin{tabular}{c|cccccc}
            \Xhline{1pt}
            & $\bot$ & $\tit{sqr}_1$ & $*_1$ & $*_2$ & $+_1$ & $+_2$ \\ 
            \Xhline{1pt}
            $\tit{sqr}$ & \\ 
            $*$  & \multicolumn{6}{c}{\multirow{2}{*}{all probabilities are}}\\ 
            $+$  & \multicolumn{6}{c}{\multirow{2}{*}{initialized to 0.20}}\\ 
            $x$  & \\ 
            $y$ &  \\ 
            \Xhline{1pt}
           \end{tabular}
    \end{subtable}
    \vspace{0.3em}

    \begin{subtable}{\textwidth}
        \centering
        \small
        \caption{The model updated after querying \toolname{GPT-4o}.}
        \label{table:model-updated}
        \begin{tabular}{c|cccccc}
            \Xhline{1pt}
            & $\bot$ & $\tit{sqr}_1$ & $*_1$ & $*_2$ & $+_1$ & $+_2$ \\ 
            \Xhline{1pt}
            $\tit{sqr}$ & \bluec{0.09} & 0.22 & 0.17 & 0.17 & 0.13 & 0.13\\ 
            $*$  & \var{0.36} & 0.22 & 0.17 & 0.17 & 0.13 & 0.13\\ 
            $+$  & 0.18 & \bluec{0.11} & \var{0.33} & \var{0.33} & 0.13 & 0.13\\ 
            $x$  & 0.18 & 0.22 & 0.17 & 0.17 & \var{0.53} & \bluec{0.07}\\ 
            $y$ & 0.18 & 0.22 & 0.17 & 0.17 & \bluec{0.07} & \var{0.53} \\ 
            \Xhline{1pt}
           \end{tabular}
    \end{subtable}
\end{table}
\end{minipage}
\end{figure*}

\autoref{fig:first-egraph} shows the saturated e-graph, which contains 10 expressions: some make progress toward the simplification, such as $x \cdot (x + y) + y \cdot (x + y)$, while others do not, such as $(y + x)^2$.
Among them, \mainname aims to extract the expression $p^*$ that makes the most progress.
It will record the rewrite chain of $p^*$ into the result, and focus on simplifying $p^*$ in the next phase.

\newcommand{\prob}[2]{\gamma(#2\,|\,#1)}

One straightforward approach to extracting $p^*$ is to scan over all programs in the e-graph, query the LLM to compare each program with the previous result, and updates the result if the LLM prefers the former.
This approach, however, is impractical because a saturated e-graph typically contains an exponential number of programs.

\mainname improves this approach by incorporating a probabilistic model to predict suitable checkpoints, and querying the LLM only with probable programs, thus reducing the search space.
The model is learned on the fly from the responses of the LLM.

\paragraph{Bigram model} \mainname utilizes a bigram model, as shown in \autoref{table:models}.
The model assigns a probability for each operator $\tit{op}$ under each context $c$, denoted as $\prob{c}{\tit{op}}$; then the probability of a program is computed as the product of the probabilities of its operators.
Except for the special context $\bot$ that denotes the topmost operator, each other context comprises two parts: the parent operator and the index of the current operator. For example, $\prob{+_1}{x}$ denotes the probability for variable $x$ to appear as the first child of $+$.
The index in the context enables our model to distinguish symmetric programs.
It can assign different probabilities to $x + y$ and $y + x$, as shown below, which must otherwise be the same without the index.
$$
\Pr[x + y] \coloneqq \prob{\bot}{+} \cdot \var{\prob{+_1}{x} \cdot \prob{+_2}{y}}
\qquad
\Pr[y + x] \coloneqq \prob{\bot}{+} \cdot \var{\prob{+_2}{x} \cdot \prob{+_1}{y}}
$$

Another advantage of this model lies in its simplicity, which allows efficient extraction from e-graphs.
Within polynomial time, we can both sample a random program and extract the most probable program from the e-graph according to the model.

\paragraph{Extraction guided by the model} 
To extract from the e-graph, \mainname starts with a default model where all operators are assigned the same probability (shown in \autoref{table:default-model}) and samples a random program as the initial result.
Then, \mainname refines the model and updates the result by iteratively querying the LLM.
In each round, it will (1) sample a new candidate program from the e-graph according to the model, (2) query the LLM to compare the previous result and the new candidate, and (3) update the model and the result according to the response. 

Suppose the initial result is $(y + x)^2$, and program $(x + y) \cdot (x + y)$ is sampled in the first iteration.
\mainname will query the LLM to decide which program makes more progress, as shown in \autoref{fig:first-query}.
Given the response that the new program makes more progress, \mainname will take this program as the new result and refine the model by enhancing the probability of the new program while weakening the probability of the previous program.
In more detail, \mainname will collect the model parameters that contribute to the probabilities of the two programs, as listed below.
\begin{align*}
\Pr[(y + x)^2] &\coloneqq \prob{\bot}{\tit{sqr}} \cdot \prob{\tit{sqr}_1}{+} \cdot \prob{+_1}{y} \cdot \prob{+_2}{x} \\
\Pr[(x + y) \cdot (x + y)]  & \coloneqq \prob{\bot}{*} \cdot \prob{*_1}{+} \cdot \prob{*_2}{+} \cdot \prob{+_1}{x}^2 \cdot \prob{+_2}{y}^2
\end{align*}
\mainname will multiply each parameter by $\alpha$ for each time it contributes to the better program, and divide it by $\alpha$ for each time it contributes to the worse one, where $\alpha > 1$ is a pre-defined constant.
For example, when $\alpha = 2$, $\prob{+_1}{x}$ will be multiplied by $4$, and $\prob{+_1}{y}$ will be divided by $2$. 

\autoref{table:model-updated} shows the refined model, where the probability sum is normalized to $1$ under each context, and we mark the enhanced parameters as \var{red} and weakened parameters as \bluec{blue}.
\mainname will use this new model to sample the next candidate program -- this process will repeat for a pre-defined number of rounds, and the final result will be taken as the checkpoint.

\smallskip 

\newcommand{\tcode}[1]{\texttt{\fontsize{10pt}{13.2pt}#1}}

\begin{figure*}
\begin{minipage}{0.55\textwidth}
    \fbox{
\begin{minipage}{0.98\textwidth}
\small 
\textbf{User:} 
I am simplifying the s-expression \tcode{(sqr (+ x y))} on a commutative ring with characteristic 2.  
Please compare the following two intermediate results, and decide which makes more progress toward simplification.  
\begin{itemize}
    \item[(a)] \tcode{(sqr (+ y x))}
    \item[(b)] \tcode{(* (+ x y) (+ y x))} 
\end{itemize}
Conclude your output with \textbf{the answer is (a)/(b)}.

\smallskip 

\textbf{Assistant}: \textbf{the answer is (b)}.
\end{minipage}}
\caption{A query to \toolname{GPT-4o} and the response.}
\label{fig:first-query}
\end{minipage}
\hfill 
\begin{minipage}{0.4\textwidth}
    \vspace{-1.5em}
    \begin{table}[H]
    \centering
        \small
        \caption{The checkpoints found by \mainname.} 
        \label{table:checkpoints}
        \renewcommand{\arraystretch}{1.2}
        \begin{tabular}{cc}
            \Xhline{1pt}
            Phase & Checkpoint \\ 
            \Xhline{1pt}
            Init & $(x+y)^2$ \\ 
            1 & $(x + y) \cdot (x + y)$ \\ 
            2 & $(x \cdot x + y \cdot x) + (x \cdot y + y \cdot y)$ \\ 
            3 & $(x \cdot x + 0) + y \cdot y$ \\ 
            4/5 & $x^2 + y^2$ \\ 
            \Xhline{1pt}
           \end{tabular}
        \end{table}
\end{minipage}
\end{figure*}

Compared to the direct approach that queries each program one by one, the key advantage of \mainname is its ability to generalize across programs with similar structures.
For example, in the refined model, the probability of $x + y$ is nearly $100$ times higher than $y + x$.
Such generalization makes \mainname focus on programs involving only $x+y$, thus immediately cutting off the program space by $70\%$ because only 3 out of the 10 programs in the e-graph satisfy this restriction.

Of course, this advantage hinges on the assumption that programs with significant progress share unique syntactic features.
We make this assumption because \citet{DBLP:journals/pacmpl/KoehlerGBGTS24} shows that human users can identify useful checkpoints via some fixed sketches, which are purely syntactic.

\paragraph{All checkpoints} In the first phase, \mainname will extract $(x + y) \cdot (x + y)$ as the checkpoint. 
Notably, \toolname{GPT-4o} does not prefer the program $x \cdot (x + y) + y \cdot (x + y)$ in the e-graph, although it is one step closer to the target expression.
This is because \toolname{GPT-4o} cannot recognize low-level rewrites -- it will directly connect $(x + y) \cdot (x + y)$ to $(x \cdot x + x \cdot y) + (y \cdot x + y \cdot y)$, skipping over five intermediate rewrites.
Fortunately, this issue will not affect the effectiveness of \mainname, because $(x + y) \cdot (x + y)$ still makes positive progress toward the target and therefore serves as a valid checkpoint.

Then, \mainname moves to the next rewrite phase, restarts equality saturation from the checkpoint, and extracts the next one from the e-graph.
This process repeats until convergence, where \mainname identifies 4 checkpoints (\autoref{table:checkpoints})  and finally returns $x^2 + y^2$, the expected result.
%!TEX root = ../paper.tex
\section{Evaluation} \label{section:evaluation}
We have implemented a prototype of \mainname and evaluated it on a constructed dataset.

\paragraph{Dataset}
We consider the problem of factorizing multi-variable polynomials, such as rewriting $x \cdot x + x \cdot y + x \cdot z + y \cdot z$ into $(x + y) \cdot (x + z)$, which can be regarded as an optimization problem for improving the efficiency of evaluating polynomials. 
This problem is challenging because it requires applying the associative and commutative laws many times, which typically leads to a combinatorial explosion in the search space.

We implement a random generator to construct tasks with various difficulties.
Our generator takes two parameters $n_d$ and $n_v$, denoting the degree of the polynomial and the number of different variables. 
It first samples $n_d$ random (non-empty) subsets of variables, then multiplies the sum of each subset, unfolds the product, and at last applies the associative and commutative laws to randomly reorganize the expression.
The following demonstrates a sample generation.
\begin{align*}
 \langle n_d = 2, n_v = 3 \rangle \xRightarrow{\text{sample}} \{y\}, \{x, y, z\} \xRightarrow{\text{multiply}} y \cdot (x + y + z) \xRightarrow{\text{unfold}} y \cdot x + y \cdot y + y \cdot z& \\ 
 \xRightarrow{\text{randomly reorganize}} z \cdot y + (x \cdot y + y \cdot y)&
\end{align*}
The goal of each task is to rewrite the last expression back to the multiplication form.

We construct our dataset by running the generator $20$ times for each combination of $n_d$ and $n_v$ within the range $[2, 5]$, thus creating tasks with a diverse range of difficulties from easy to hard.

\paragraph{Baselines}

We compare \mainname with two baseline solvers.
\begin{itemize}
\item \toolname{DirectES} denotes direct equality saturation. It creates an e-graph for the initial expression and saturates the e-graph via \toolname{egg} until a multiplication form is reached.
\item \toolname{DirectLLM} directly generates the rewrite chain by LLMs. It queries the LLM with the task description and the rewrite system, then verifies the rewrite chain in the response.
\end{itemize}

\paragraph{Configuration}Our experiments are conducted on Intel Xeon Gold 6230 2.1GHz 20-Core Processor, with a timeout of 150 seconds and a memory limit of 24GB per task. 

In each phase of \mainname, we set a timeout of 5 seconds for equality saturation, limit the number of rounds for refining the bigram model to $10$, and set the constant $\alpha$ to $1.5$. 
Besides, we use \toolname{GPT-4o} as the backend LLM for both \mainname and \toolname{DirectLLM}.

\begin{figure*}
    \hfill
    \begin{subfigure}[b]{0.31\textwidth}
        \centering
        \includegraphics[height=12.5em]{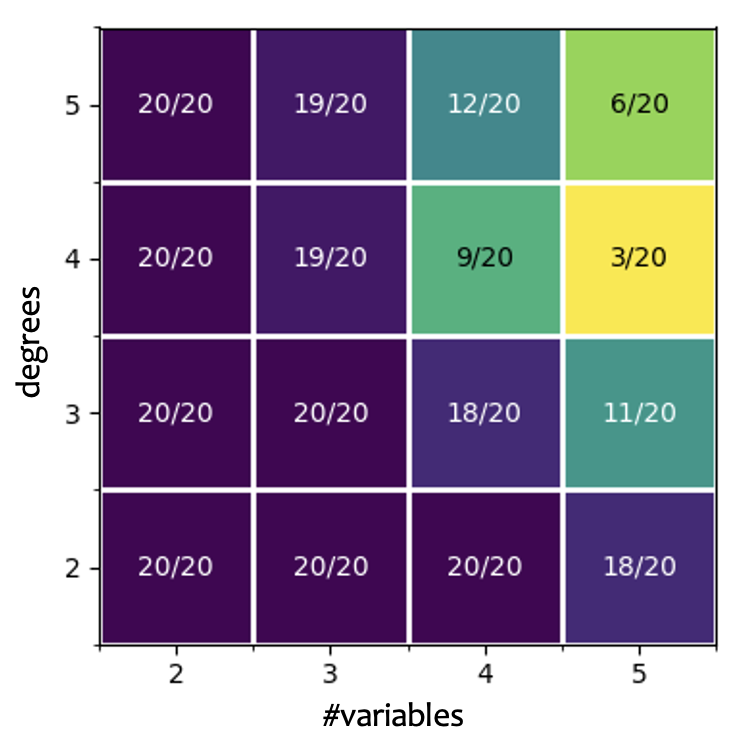}
        \caption{\mainname solves 255/320 tasks.}
    \end{subfigure}
    \hfill
    \begin{subfigure}[b]{0.31\textwidth}
        \centering
        \includegraphics[height=12.5em]{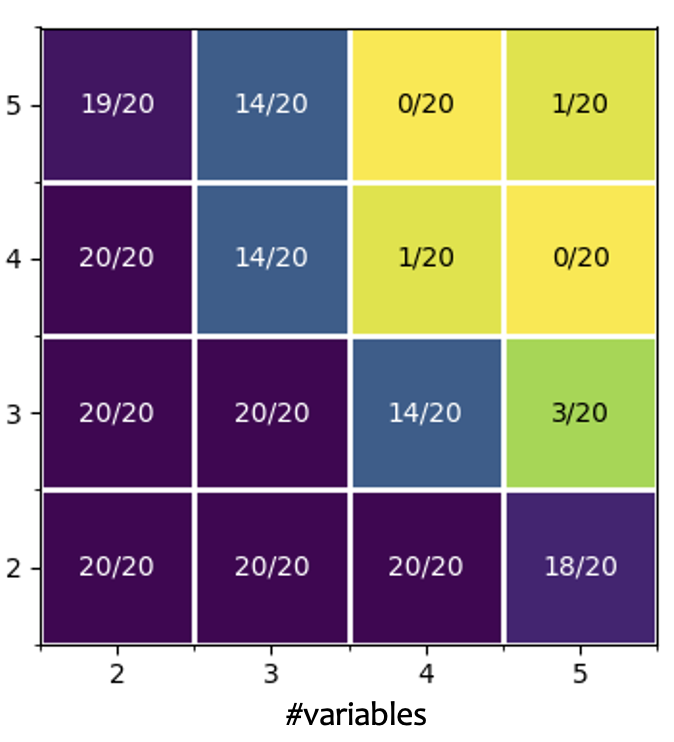}
        \caption{\toolname{DirectES} solves 204/320 tasks.}
    \end{subfigure}
    \hfill
    \begin{subfigure}[b]{0.34\textwidth}
        \centering
        \includegraphics[height=12.5em]{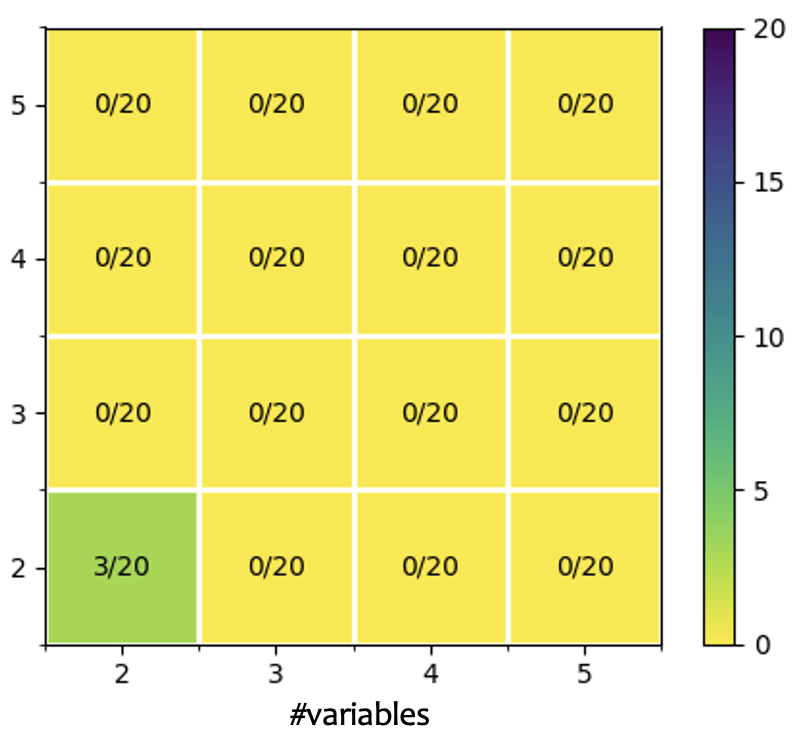}
        \caption{\toolname{DirectLLM} solves 3/320 tasks.\ \ \ }
    \end{subfigure}
    \hfill\ 
    \caption{The results of our evaluation. In these figures, each row denotes a different degree of polynomials, and each column denotes a different number of variables -- the factorization task becomes more and more challenging from lower left to upper right. In each cell, we report the number of solved tasks in the corresponding subset, where a deeper color denotes a better performance.}
    \label{figure:evaluation-results}
\end{figure*}

\paragraph{Evaluation results} We run \mainname and the two baseline solvers on all tasks in our dataset.
The results are summarized in \autoref{figure:evaluation-results}.
They demonstrate that \mainname can solve most task in the dataset (255 out of 320) and significantly outperforms both baseline solvers.
\begin{itemize}
        \item Compared with \toolname{DirectES}, \mainname has a clear advantage on challenging tasks -- it solves 30 tasks when the degree and the number of variables are both no smaller than $4$, while \toolname{DirectES} can only solve $2$. This is because \mainname performs equality saturation in phases, thus alleviating the combinatorial explosion problem in \toolname{DirectES}.
        \item \toolname{DirectLLM} fails to solve almost all tasks because \toolname{GPT-4o} can hardly maintain the low-level rewrite chain -- it frequently makes mistakes such as skipping steps and applying unavailable rules.
        In contrast, \mainname addresses this issue by incorporating e-graphs to handle all low-level rewrites.
\end{itemize}
%!TEX root = ../paper.tex
\section{Conclusion and Future Work} \label{section:conclusion}
In this paper, we study the correctness issue of LLMs and propose a novel approach \mainname, which bridges the gap between LLMs and formal rewrite systems via e-graphs.
We evaluate \mainname on a constructed dataset for polynomial factorization and preliminarily verify its effectiveness.

In the next step, we will instantiate \mainname in more practical domains, such as optimizing CAD programs~\cite{DBLP:conf/pldi/YaghmazadehKDC16} and array programs~\cite{DBLP:journals/pacmpl/HagedornLKQGS20}, to see whether \mainname can effectively handle real-world problems. 
Besides, we will also investigate the reasons why \mainname fails on some tasks in our dataset and work to improve it accordingly.  

\interlinepenalty=10000
\bibliography{ref} 

\end{document}